\documentclass[aps,prl,floats,twocolumn,showpacs,superscriptaddress]{revtex4}
\usepackage{epsfig}

\newcommand{\avk}{\left< k \right>}
\newcommand{\fluck}{\left< k^2 \right>}
\newcommand{\condP}{P(k' \, \vert \, k)}

\begin{document}
 
\title{Epidemic spreading in correlated complex networks} 

\author{Mari{\'a}n Bogu{\~n}{\'a}} 

\affiliation{Departament de F{\'\i}sica Fonamental, Universitat de
  Barcelona, Av. Diagonal 647, 08028 Barcelona, Spain}

\author{Romualdo Pastor-Satorras}

\affiliation{Departament de F{\'\i}sica i Enginyeria Nuclear, Universitat
  Polit{\`e}cnica de Catalunya, Campus Nord B4, 08034 Barcelona, Spain}

\date{\today} 

\begin{abstract} 
  We study a dynamical model of epidemic spreading on complex networks
  in which there are explicit correlations among the node's
  connectivities. For the case of Markovian complex networks, showing
  only correlations between pairs of nodes, we find an epidemic
  threshold inversely proportional to the largest eigenvalue of the
  connectivity matrix that gives the average number of links that from
  a node with connectivity $k$ go to nodes with connectivity $k'$.
  Numerical simulations on a correlated growing network model provide
  support for our conclusions.
\end{abstract} 
 

\maketitle 

Statistical physics has witnessed in recent years a renewed interest
in graph theory due to the discovery that many natural and artificial
systems can be described in terms of complex networks, in which the
nodes represent typical units, and the links represent the
interactions between pairs of units \cite{barabasi01,dorogorev}. The
term complex network has been coined to refer to networks that
typically exhibit two distinct properties: (i) A \textit{scale-free}
(SF) connectivity distribution. If we define the connectivity
distribution, $P(k)$, as the probability that a node is connected to
$k$ other nodes, then SF networks are characterized by a power-law
behavior $P(k) \sim k^{-\gamma}$, where $\gamma$ is a characteristic exponent.
This property renders classical models of random graphs
\cite{erdos60}, described by an exponentially bounded connectivity,
inappropriate for the description of many real networks. (ii) The
\textit{small-world} property \cite{watts98}, that is defined by an
average path length---average distance between any pair of
nodes---increasing very slowly (usually logarithmically) with the
network size $N$.

Apart from the empirical characterization of real complex networks and
the development of new models, accounting for the observed properties,
the interest in this field has been also driven by the discovery of
the profound and subtle effects that the connectivity has on the
behavior of dynamical systems defined on top of complex networks.
These effects are particularly interesting in the study of disease
transmission in SF networks, relevant for the understanding of the
spreading of computer viruses \cite{virusreview} and sexually
transmitted diseases \cite{amaral01}.  Indeed, it was first noted in
Ref.~\cite{pv01a} that in \textit{uncorrelated} SF networks with a
connectivity exponent $\gamma \leq 3$, epidemic processes do not posses an
epidemic threshold below which diseases cannot produce a macroscopic
epidemic outbreak or the development of an endemic state.  This
feature, observed in several epidemic models
\cite{pv01a,pv01b,lloydsir,moreno,newman02}, is deeply rooted in the
presence of very large connectivity fluctuations in infinite SF
networks.

The study of epidemic spreading in uncorrelated complex networks (that
is, in graphs in which the connectivity of any node is independent of
the connectivity of its neighbors) has been proved to be extremely
successful, providing, for instance, the first satisfactory
explanation of the long-standing problem of the generalized low
prevalence of computer viruses without assuming any global tuning of
microscopic parameters \cite{pv01a,white98}.  Nevertheless, it
represents a first approximation to real networks, that neglects the
possibility of correlations in the connectivity of the nodes. The
existence and importance of connectivity correlations has been
recently pointed out in the literature.  In fact, it has become clear
that these correlations are critical in the understanding of the
hierarchical structure of the Internet \cite{alexei,alexei2,goh01b}.
On the other hand, some growing network models have been proposed
\cite{callaway}, in which correlations are spontaneously generated and
have important effects in the percolation transition.

In this paper we shall present a study of epidemic spreading in
complex random networks in which there are explicit correlations among
the node's connectivities. We will consider in particular the subset
of undirected \textit{Markovian} random networks, that are completely
defined by their connectivity distribution $P(k)$ and the conditional
probability $\condP$ that a node of connectivity $k$ is connected to a
node of connectivity $k'$. These two functions can have any form
(including SF behavior), and are assumed to be normalized ($\sum_k P(k)
= \sum_{k'} \condP = 1$) and restricted by the connectivity detailed
balance condition 
\begin{equation}
  k \condP P(k) = k' P(k \, \vert \, k') P(k') \equiv \avk P(k, k'),
  \label{eq:1}
\end{equation}
where the symmetric function $(2 - \delta_{k  k'}) P(k, k')$ is the joint
probability that two nodes of connectivity $k$ and $k'$ are connected.
The Markovian nature of this class of networks implies that all higher
order correlations can be expressed as a function of $P(k)$ and
$\condP$, allowing an exact treatment of epidemic models at the
mean-field (MF) level. It is worth noticing, however, that a more
detailed description, in terms of a Langevin equation (to be reported
elsewhere \cite{marian}), yields exactly the same results, confirming
the accuracy of the MF description.  In this framework, the
topologically relevant magnitude is the connectivity matrix $C_{k k'}
= k \condP$, that measures the average number of links that go from a
node with connectivity $k$ to nodes with connectivity $k'$.  We will
show that the epidemic threshold is related to the largest eigenvalue
of this matrix. Extensive numerical simulations on a random correlated
network model \cite{callaway} confirm the predictions of the present
analysis.  During the completion of this work we became aware of two
recent preprints \cite{assortative,structured} in which it is also
highlighted the general role of correlations in spreading and
percolation in complex networks.

In order to study the effects of connectivity correlations in epidemic
spreading, we will focus in the standard
susceptible-infected-susceptible (SIS) model \cite{epidemics}. All the
results, however, can be easily extended to the more general
susceptible-infected-removed-susceptible (SIRS) model \cite{marian}.
In the SIS model each node in the network represents an individual,
and each link represents a connection along which the infection can
propagate. Susceptible (healthy) nodes become infected with
probability $\nu$ if they are connected to one or more infected nodes.
On the other hand, infected nodes recover spontaneously with
probability $\delta$.  The ratio of this two rates defines an effective
spreading rate $\lambda=\nu / \delta$ (without lack of generality, we set
$\delta=1$). For homogeneous networks, in which each node has more or less
the same number of connections, $k \simeq \avk$, a general result states
the existence of a finite epidemic threshold, separating an infected
(endemic) phase, with a finite average density of infected
individuals, from a healthy phase, in which the infection dies out
exponentially fast. In terms of the average density of infected
individuals $\rho(t)$ (the prevalence) we can describe the SIS model in
homogeneous networks at a MF level by the following rate equation
\cite{pv01b}
\begin{equation}
  \partial_t \rho(t) = -\rho(t) +\lambda \avk \rho(t) \left[ 1-\rho(t) \right].
\label{eq:ws}
\end{equation}
In this equation we have neglected higher order terms, since we are
interested in the onset of the endemic state, close to the point
$\rho(t) \sim 0$. Also, we have neglected correlations among nodes.  That
is, the probability of infection of a new node---the second term in
Eq.~(\ref{eq:ws})---is proportional to the infection rate $\lambda$, to the
probability that a node is healthy, $1-\rho(t)$, and to the probability
that a link in a healthy node points to an infected node.  This last
quantity, assuming the \textit{homogeneous mixing hypothesis}
\cite{anderson92}, is approximated for homogeneous networks as $\avk
\rho(t)$, i.e. proportional to the average number of connection and to
the density of infected individuals, and independent of the
connectivity.  From Eq.~(\ref{eq:ws}) it can be proved the existence
of an epidemic threshold $\lambda_c= \avk^{-1}$ \cite{epidemics}, such that
$\rho = 0$ if $\lambda< \lambda_c$, while $\rho \sim (\lambda-\lambda_c)$ if $\lambda\geq \lambda_c$.  In
this context, it is easy to recognize that the SIS model is a
generalization of the contact process model, widely studied as the
paradigmatic example of an absorbing-state phase transition to a
unique absorbing state \cite{marro99}.

For general complex networks, in which large connectivity fluctuations
and correlations might be allowed, we must relax the homogeneous
hypothesis made in writing Eq.~(\ref{eq:ws}) and work instead with the
relative density $\rho_k(t)$ of infected nodes with given connectivity
$k$; i.e.  the probability that a node with $k$ links is infected.
Following Refs.~\cite{pv01a,pv01b}, the rate equation for $\rho_k(t)$
can be written as
\begin{equation}
  \frac{ d \rho_k(t)}{d t} = 
  -\rho_k(t) +\lambda k \left[1-\rho_k(t) \right]  \Theta_k(t).  
\label{mfk}
\end{equation}
In this case, the creation term is proportional to the spreading rate
$\lambda$, the density of healthy sites $1-\rho_k(t)$, the connectivity $k$,
and the variable $\Theta_k(t)$, that stands for the probability that a
link emanating from a node of connectivity $k$ points to an infected
site.  In the case of an uncorrelated random network, considered in
Refs.~\cite{pv01a,pv01b}, the probability that a link points to a node
of connectivity $k'$ is \textit{independent} of the connectivity $k$
of the node from which the link is emanating.  Therefore, in this case
$\Theta_k=\Theta^{\rm nc}$ is independent of $k$ and can be written as
\begin{equation}
  \Theta^{\rm nc} =\frac{1}{\avk}\sum_{k'} k' P(k')\rho_{k'}(t),
  \label{first}
\end{equation}
since the probability that a node is pointing to a node of
connectivity $k'$ is proportional to $k' P(k')$. Substituting the
expression (\ref{first}) into Eq.~(\ref{mfk}), one can solve for the
steady state solution and find the existence of an epidemic threshold
$\lambda_c$, below which there are no solutions with a nonzero value of
$\Theta^{\rm nc}$. The expression of the epidemic threshold for
uncorrelated random networks is
\begin{equation}
  \lambda_c^{\rm nc} =\frac{\avk}{\fluck}.
  \label{eq:2}
\end{equation}
For infinite SF networks with $\gamma \leq 3$, we have $\fluck = \infty$, and
correspondingly $\lambda_c^{\rm nc} = 0$.  Finally, from the solution of
$\rho_k$, one can compute the total prevalence $\rho$ using the relation
$\rho=\sum_kP(k)\rho_k$.

For a general network in which the connectivities of the nodes are
correlated, the above formalism is not correct, since we are not
considering the effect of the connectivity $k$ into the expression for
$\Theta_k$. This effect can be taken into account, however, for Markovian
networks, whose correlations are completely defined by the conditional
probability $\condP$.  In this case, it is easy to realize that the
correct factor $\Theta_k$ can be written as
\begin{equation}
  \Theta_k(t) = \sum_{k'} \condP \rho_{k'}(t),
  \label{eq:3}
\end{equation}
that is, the probability that a link in a node of connectivity $k$ is
pointing to an infected node is proportional to the probability that
any link points to a node with connectivity $k'$, times the
probability that this node is infected, $\rho_{k'}(t)$, averaged over
all the nodes connected to the original node. Eqs.~(\ref{mfk})
and~(\ref{eq:3}) define together the MF equation describing the SIS
model on Markovian complex networks,
\begin{equation}
  \frac{ d \rho_k(t)}{d t} = 
  -\rho_k(t) +\lambda k \left[1-\rho_k(t) \right] \sum_{k'} \condP \rho_{k'}(t). 
  \label{generalized}
\end{equation}

The exact solution of Eq.~(\ref{generalized}) can be difficult to
find, depending on the particular form of $\condP$. However, it is
possible to extract the value of the epidemic threshold by analyzing
the stability of the steady-state solutions. Of course, the healthy
state $\rho_k=0$ is one solution. For small $\rho_k$, we can linearize
Eq.~(\ref{generalized}), getting
\begin{equation}
  \frac{ d \rho_k(t)}{d t} \simeq \sum_{k'} L_{k  k'} \rho_{k'}(t).
\end{equation}
In the previous equation we have defined the Jacobian matrix
$\mathbf{L}=\{ L_{k  k'} \}$ by
\begin{equation}
  L_{k  k'} = -\delta_{k  k'} + \lambda k \condP,
\end{equation}
where $\delta_{k k'}$ is the Kronecker delta function. The solution
$\rho_k=0$ will be unstable if there exists at least one positive
eigenvalue of the Jacobian matrix $\mathbf{L}$.  Let us consider the
\textit{connectivity matrix} $\mathbf{C}$, defined by $C_{k k'} = k
\condP$. Using the symmetry condition Eq.~(\ref{eq:1}), it is easy to
check that if $v_k$ is an eigenvector of $\mathbf{C}$, with eigenvalue
$\Lambda$, then $P(k) v_k$ is an eigenvector of the transposed matrix
$\mathbf{C}^T$ with the same eigenvalue. From here it follows
immediately that all the eigenvalues of $\mathbf{C}$ are real.
Let $\Lambda_m$ be the
largest eigenvalue of  $\mathbf{C}$.
Then, the origin will be unstable whenever $- 1 + \lambda \Lambda_m >0$, which
defines an epidemic threshold
\begin{equation}
  \lambda_c = \frac{1}{ \Lambda_m},
  \label{eq:4}
\end{equation}
above which the solution $\rho_k=0$ is unstable, and another nonzero
solution takes over as the actual steady-state---the endemic state.

It is instructive to see how this general formalism recovers previous
results \cite{pv01a,pv01b}, implicitly obtained for random
uncorrelated networks. For any random network, in which there are no
correlations among the connectivities of the nodes, we have that the
connectivity matrix is given by $C_{k k'}^{\rm nc} = k P(k' / k) \equiv k
k' P(k')/ \avk$, since the probability that a link points to a node of
connectivity $k'$ is proportional to $k' P(k')$. It is easy to check
that the matrix $\{ C_{k' k}^{\rm nc} \} $ has unique eigenvalue
$\Lambda_m^{\rm nc} = \fluck / \avk$, corresponding to the eigenvector
$v_k^{\rm nc} = k$, from where we recover the now established result
Eq.~(\ref{eq:2}).

In order to check the theoretical prediction Eq.~(\ref{eq:4}), we have
performed numerical simulations of the network model proposed by
Callaway \textit{et al.} \cite{callaway}. This model is very simply
defined: each time step we add a new node, and, with probability $\delta$,
two nodes are randomly selected and joined with a link. This recipe
yields a random network with a exponential connectivity distribution,
$P(k)=(2 \delta)^k / (1 + 2 \delta)^{k+1}$, which shows correlations among the
connectivities of the nodes. These correlations can be analytically
computed by means of the joint probability $P(k, k')$, which fulfills
the recursion relation \cite{callaway}
\begin{eqnarray*}
  P(k,k')&=&\frac{2\delta}{1+4\delta} [ P(k-1,k')+P(k,k'-1) ] \\
  &+& \frac{P(k-1)P(k'-1)}{1+4\delta}.
\end{eqnarray*}
Even though it is not possible to solve for $P(k,k')$ in a closed form,
we can numerically estimate the largest eigenvalue of the connectivity
matrix $\mathbf{C}$ by generating a finite matrix $P(k, k')$ and using
the relation (\ref{eq:1}) to find $\condP$. The correctedness of this
numerical estimate is ensured by the exponential decay of $P(k, k')$
for large values of $k$ and $k'$, as can be seen from the recursion
relation. The estimate of the largest eigenvalue of the connectivity
matrix for Callaway's model is $\Lambda_m \approx 6.47656$, which yields an
epidemic threshold $\lambda_c \approx 0.15$. This value is to be compared with
the prediction for an uncorrelated network, $\lambda_c^{\rm nc} = \avk /
\fluck = 0.20$.

\begin{figure}[t]
  \epsfig{file=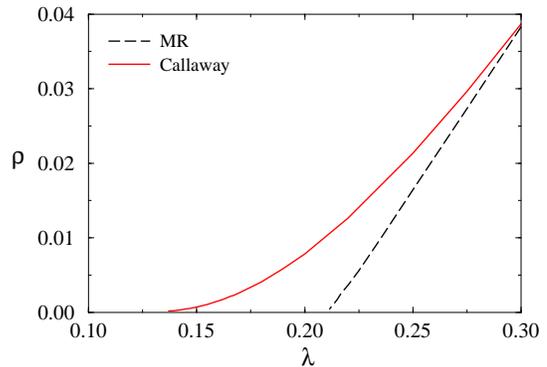, width=7cm}
    
  \caption{Total prevalence $\rho$ for the SIS model in Callaway's
    network and in a MR uncorrelated network with the same
    connectivity distribution.}
  \label{fig:rho}
\end{figure}

\begin{figure}[t]
  \epsfig{file=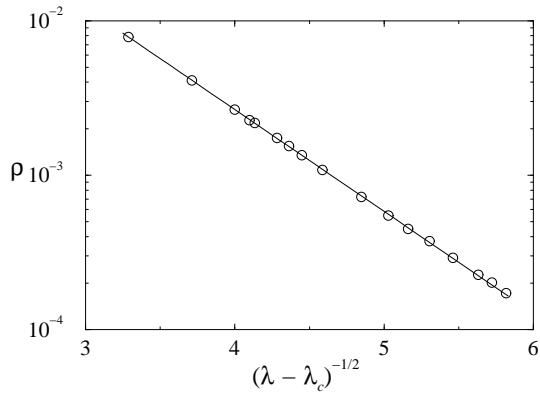, width=7cm}
    
  \caption{Fit of the prevalence for the SIS model in Callaway's
    network to the form $\rho(\lambda) \sim \exp [-\alpha (\lambda-\lambda_c)^{-1/2}]$. The
    fit yields the values $\alpha=1.52\pm0.02$ and $\lambda_c = 0.11\pm0.02$.}
  \label{fig:fit}
\end{figure}

Figure~\ref{fig:rho} shows the results of numerical simulations of the
SIS model on Callaway's network, as well as on random networks with
the same connectivity distribution, generated using the Molloy and
Reed (MR) algorithm \cite{molloy95,molloy98}. The MR algorithm
generates a random network with a prescribed connectivity distribution
and no correlations among nodes, and thus it is expected to yield an
epidemic threshold given by $\lambda_c^{\rm nc}$.  Simulations were
performed for a fixed value $\delta=1$ in networks of size up to $N=10^7$,
averaging over at least $100$ different starting configurations,
performed on at least $10$ different realizations of the network.
Figure~\ref{fig:rho} depicts the steady state prevalence as a function
of the spreading rate $\lambda$. For the MR network, the function $\rho(\lambda)$
shows a clear linear behavior. The epidemic threshold estimated from a
least-squares fitting is $\lambda_c=0.21\pm0.01$, in excellent agreement with
the prediction for an uncorrelated network. On the other hand,
Callaway's network exhibits a very different behavior, which might be
compatible with the presence of a transition of infinite order
\cite{marro99}. In fact, Fig.~\ref{fig:rho} is reminiscent of the
behavior found in Ref.~\cite{callaway} for the giant component of the
network as a function of the parameter $\delta$. In that work, the size of
the giant component was fitted to an stretched exponential with
exponent $1/2$. Guided by this intuition, in Fig.~\ref{fig:fit} we
perform a fit of the stationary prevalence in Callaway's model to the
form $\rho(\lambda) \sim \exp [-\alpha (\lambda-\lambda_c)^{-1/2}]$ \cite{callaway}. The fit
yields a prefactor $\alpha=1.52\pm0.02$, and an epidemic threshold $\lambda_c =
0.11\pm0.02$, smaller by a factor $2$ than the value corresponding to an
uncorrelated random network, and in quite good agreement with the
prediction from the largest eigenvalue of the connectivity matrix.

In summary, we have shown that, in the presence of correlations, the
epidemic threshold in complex networks is determined by the
connectivity matrix $\mathbf{C}$, and not by the connectivity
distribution $P(k)$, as happens in uncorrelated networks. This fact
implies that the previously predicted null epidemic threshold for SF
networks with $\gamma\leq 3$ might be shifted in correlated graphs,
attaining a positive value depending on the nature of the correlations
as given by the connectivity matrix. At this respect, it might be
surprising that some complex networks, such as the Barab{\'a}si-Albert
(BA) graph \cite{barab99}, are exactly described at the uncorrelated
level given by Eq.~(\ref{mfk}) with $\Theta_k$ independent of $k$
\cite{pv01a,pv01b}. This fact must be taken as an evidence of the lack
of correlations in the BA model; lack that, on the other hand, has
been already checked numerically in Ref.~\cite{alexei2}.  The
formalism presented in this paper represents a refinement over
previous works because it includes the effects of correlations between
pairs of nodes and, in this sense, it is exact for Markovian networks.
Real networks, such as the Internet, however, posses a more complex
correlation structure.  Our formalism will provide an improved
approximation to epidemic dynamics in these cases, but it still
remains the task of ascertaining the effects of higher order
correlations. Further work is necessary in this direction.

\begin{acknowledgments}
  This work has been partially supported by the European Commission -
  Fet Open project COSIN IST-2001-33555. R.P.-S. acknowledges
  financial support from the Ministerio de Ciencia y Tecnolog{\'\i}a
  (Spain). We thank A. Vespignani for helpful comments and
  discussions.
\end{acknowledgments}

\end{document}